# Sodium and oxygen nonstoichiometry, and thermoelectric properties of $Na_xCoO_{2+\delta}$

T. Motohashi, M. Karppinen, and H. Yamauchi

*Materials and Structures Laboratory, Tokyo Institute of Technology, Yokohama 226-8503, Japan*

**Abstract**

The sodium-cobalt oxide $Na_xCoO_{2+\delta}$ is a promising candidate for thermoelectric applications since it may possess simultaneously large thermoelectric power and low resistivity. It is difficult to understand the simultaneous appearance of the two properties within the framework of conventional one-electron models. It has been suggested that the spin state of cobalt ions plays a crucially important role in enhancing the thermoelectric power. In order to study the relationship between the cobalt spin state and thermoelectric properties, $Na_xCoO_{2+\delta}$ samples with precisely controlled sodium and oxygen contents are indispensable. Recently, we established an unconventional sample-synthesis method that enables us to precisely control the Na content in $Na_xCoO_{2+\delta}$ samples. In this article, we discuss the effects of sodium and oxygen nonstoichiometry on the thermoelectric properties of $Na_xCoO_{2+\delta}$.

## 1. Introduction

Since the discovery of high-$T_c$ superconducting copper oxides, tremendous amount of research works have been done on various oxides of $3d$ transition metals to look for novel functions due to the strong electron-correlation effects. In 1997, Terasaki *et al*. [1] reported that the sodium-cobalt oxide, $Na_xCoO_{2+\delta}$ ($x \approx 0.5$, i.e. $NaCo_2O_4$), showed large thermoelectric power and low resistivity simultaneously. This sodium-cobalt oxide, therefore, is considered to be a promising candidate for thermoelectric applications at elevated temperatures because of its high chemical stability and non-toxicity as compared to representative thermoelectric materials such as $Bi_2Te_3$ and $PbTe$ [2]. The large magnitude of thermoelectric power, i.e. $> 50$ $\mu$V / K at 300 K, of $Na_xCoO_{2+\delta}$ is difficult to be understood within the framework of conventional one electron models [3].

It has been suggested that the spin state of cobalt ions may play the most essential role in facilitating the high thermoelectric power in $Na_xCoO_{2+\delta}$ [3].

The present authors have carefully studied the relationship between the chemical composition and physical properties of $Na_xCoO_{2+\delta}$ in order to clarify the effect of the spin state of cobalt on the thermoelectric characteristics. The crystal structure of $Na_xCoO_{2+\delta}$ consists of a single atomic layer of Na ions, being sandwiched by two-dimensional $CoO_2$ layers, i.e. an Na ion layer intercalated between adjacent $CoO_2$ layers. This structure is known to bear a wide range of Na nonstoichiometry within $0 \leq x \leq 1.0$ [4]. However, the nonstoichiometry of the phase in terms of the oxygen content is not well understood. In this article, we show that in $Na_xCoO_{2+\delta}$, (1) the sodium content can be precisely controlled within 0.60-0.65 $\leq x \leq$ 0.75-0.80 by adjusting the starting Na composition, (2) the oxygen nonstoichiometry is nearly negligible, and (3) both electrical resistivity and thermoelectric power are strongly dependent on the Na content, $x$.

## 2. Results and Discussion

*2.1 Establishment of a "rapid heat-up" technique*

First, we studied the phase formation employing a thermobalance (MAC Science, TG/DTA 2000S) in order to establish the precise control of Na nonstoichiometry for $Na_xCoO_{2+\delta}$. It had been recognized that the control of Na content is difficult for $Na_xCoO_{2+\delta}$ because Na evaporation is significant during the conventional synthesis procedure [5]. A powder mixture of $Na_2CO_3$ and $Co_3O_4$ with a molar ratio of Na:Co = 0.55:1.0, i.e. $x$ = 0.55, was heated in flowing $O_2$ gas with a heating program as shown in the inset of Fig. 1. A representative thermogravimetric (TG) curve obtained for the raw material mixture is shown in Fig. 1. From this TG analysis, it was found that the weight loss between 500 and 750°C is too large (~7.5%) to be explained by decarbonation of the raw materials (5.7%) only, suggesting that at temperatures lower than ~ 600°C where the chemical reaction between the raw materials may not be active yet the observed weight loss is due to Na evaporation.

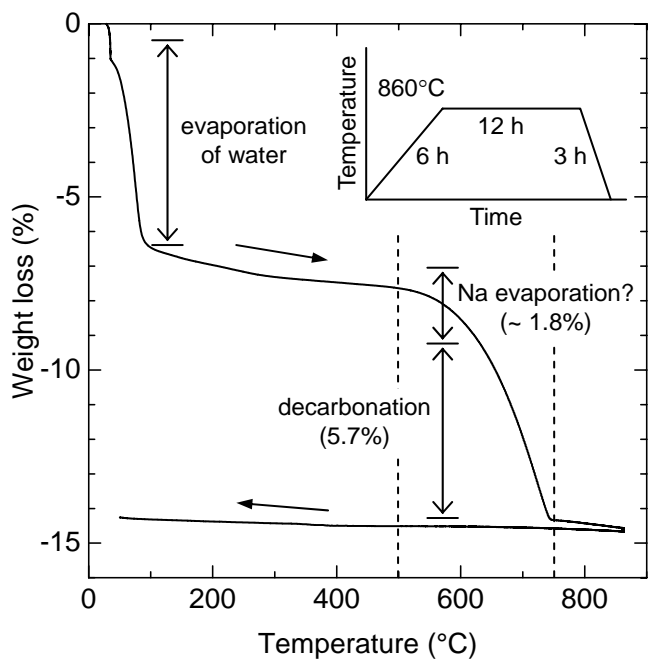

FIG. 1. Thermogravimetric curve for the phase-formation process of $Na_{0.55}CoO_{2+\delta}$ in flowing $O_2$ gas. The mass of the sample was 54.6 mg. The inset represents the heating program employed.

Based on the findings of the TG analysis on the phase formation, we employed a "rapid heat-up" technique to fire the raw materials to obtain $Na_x$-$CoO_{2+\delta}$ samples with precisely controlled Na contents by avoiding Na evaporation [6]. In this technique, a powder mixture of $Na_2CO_3$ and $Co_3O_4$ of an appropriate ratio is directly placed in a furnace preheated at 750°C to instantaneously start chemical reaction between the raw materials. The Na contents determined by ICP analysis for samples obtained by this technique were found to be in excellent agreement with the nominal ones [6]. We also confirmed that a large amount of Na is inevitably lost if the conventional procedure is employed for the preparation of $Na_xCoO_{2+\delta}$ samples.

## 2.2 Sodium nonstoichiometry in the $\gamma$-$Na_xCoO_{2+\delta}$ phase

Utilizing the rapid heat-up technique, we were able to study the Na nonstoichiometry of the $\gamma$-$Na_xCoO_{2+\delta}$ phase [7]. Samples with various Na contents were synthesized using this technique: powder mixtures of $Na_2CO_3$ and $Co_3O_4$ with molar ratios of Na:Co = (0.55 ~ 0.85) : 1.0, i.e. $x$ = 0.55 ~ 0.85, were directly placed in a furnace preheated at 750°C and fired for 12 h in air.

Figure 2 shows x-ray powder diffraction patterns for the $x$ = 0.55 ~ 0.85 samples. Although the main phase was the hexagonal $\gamma$-$Na_xCoO_{2+\delta}$ phase [4,8] for all the samples examined, some of them contained impurity phases. For the $x$ = 0.55 and 0.60 samples, slight traces of unreacted $Co_3O_4$ were seen. For the Na-rich $x$ = 0.80 and 0.85 samples, on the other hand, small peaks from unreacted $Na_2CO_3$ appeared in the diffraction patterns. It was thus concluded that the single-phase region for the $\gamma$-$Na_xCoO_{2+\delta}$ phase is: 0.60-0.65 $\leq$ $x$ $\leq$ 0.75-0.80, indicating that the stoichiometric $NaCo_2O_4$ (i.e. $Na_{0.50}CoO_2$) does not exist.

Previously, $Na_xCoO_{2+\delta}$ samples have occasionally been synthesized with nominal ratios of Na:Co = (0.55 ~ 0.60) : 1.0 to aim at the stoichiometric $NaCo_2O_4$ phase, where excess Na in the raw material mixture was assumed to compensate the loss during the synthesis procedure [2,9,10]. Moreover, most of published theoretical analyses have been performed assuming $NaCo_2O_4$ stoichiometry [11,12]. Our results strongly suggest that single-phase samples cannot be obtained for the nominal composition of $NaCo_2O_4$, even though Na evaporation is completely prevented. Thus, Na nonstoichiometry of $Na_xCoO_{2+\delta}$ should carefully be concerned with future investigations.

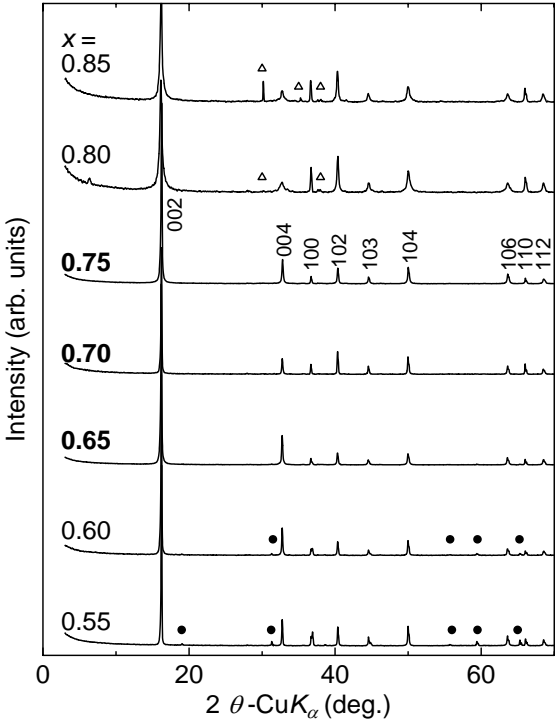

FIG. 2. X-ray powder diffraction patterns for the $x = 0.55 - 0.85$ samples prepared by the rapid heat-up technique. Peaks from unreacted $Co_3O_4$ and $Na_2CO_3$ are marked with solid circles and open triangles, respectively. The indices are for the hexagonal $\gamma$-$Na_xCoO_{2+\delta}$ phase.

## 2.3 Oxygen nonstoichiometry

Oxygen nonstoichiometry of our $Na_xCoO_{2+\delta}$ samples was studied using a wet-chemical technique [7]. Bulk samples with $x = 0.75$ were made with the rapid heat-up technique. Two types of samples with different heat treatments were prepared from the same calcined powder; the samples of type I were quenched from 900°C (named "Q"), and the samples of type II were slow-cooled from 900°C to room temperature for a period of 6 h (named "SC"). For both sample types, oxygen content $\delta$ was determined by a cerimetric titration. In the titration procedure, the sample was dissolved in an acidic solution containing a known amount of excess $Fe^{2+}$ which reduced high-valent cobalt in the sample, i.e. $Co^{3+}$ and $Co^{4+}$, and then the remaining amount of $Fe^{2+}$ was titrated by $Ce^{4+}$. When calculating the value of $\delta$ from the titration data, Na content was set to the nominal value, $x = 0.75$. A more detailed description of the experimental procedure is given elsewhere [7].

Table 1 shows the value obtained for $\delta$ and the average valence of cobalt in $Na_{0.75}CoO_{2+\delta}$. For both "Q" and "SC" types of samples the $\delta$ value was nearly zero, indicating that the oxygen nonstoichiometry is negligible. From the titration results, we may conclude that an unusually high valence state of cobalt of *ca*. +3.3 exists in $Na_xCoO_{2+\delta}$. It should be noted, however, that this value is lower than the valence state assumed for cobalt in the stoichiometric "$NaCo_2O_4$" compound.

TABLE 1. Oxygen content $\delta$ and average valence of cobalt for $Na_{0.75}CoO_{2+\delta}$ samples by the cerimetric titration. Three and two parallel batches of the quenched (Q) and slow-cooled (SC) samples, respectively, were analyzed. Furthermore, for each sample the titration was performed two or three times.

| Sample | $\delta$ value | Average valence of cobalt |
|---|---|---|
| Q (#1) | 0.038 | 3.32 |
| | 0.037 | |
| | 0.038 | |
| | Av: 0.037 | |
| Q (#2) | 0.035 | 3.32 |
| | 0.037 | |
| | Av: 0.036 | |
| Q (#3) | 0.021 | 3.29 |
| | 0.021 | |
| | 0.022 | |
| | Av: 0.021 | |
| SC (#1) | 0.019 | 3.29 |
| | 0.017 | |
| | Av: 0.018 | |
| SC (#2) | 0.020 | 3.29 |
| | 0.020 | |
| | 0.020 | |
| | Av: 0.020 | |

## 2.2 Simultaneously enhanced S and reduced $\rho$ in Na-rich samples

Electrical resistivity ($\rho$) and thermoelectric power ($S$) measurements were performed on samples with three different Na contents, i.e. $x = 0.55$, 0.65 and 0.75. Bulk specimens for transport measurements were prepared by pelletizing the sample powder prepared by our rapid heat-up technique including firing at 900°C for 12 h. Details of the experimental procedure are given elsewhere [6]. Figure 3(a) shows temperature dependences of $\rho$ for the $x = 0.55$, 0.65, and 0.75 samples from room temperature down to 5 K. All the samples show metallic behavior (i.e. $d\rho / dT > 0$) in the whole temperature range studied. It can be seen that the magnitude of $\rho$ decreases with increasing $x$, and an anomaly temperature ($T_t$) seen in the resistivity curves also decreases with $x$. These systematic changes support the conclusion that the Na content is precisely controlled in our rapid heat-up samples.

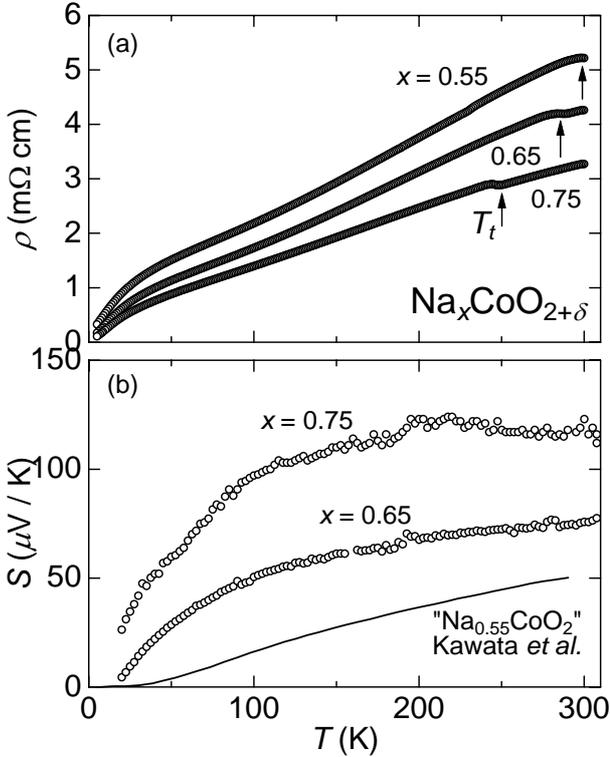

FIG. 3. Temperature dependences of $\rho$ and $S$ of $Na_xCoO_{2+\delta}$ samples. The solid curve in Fig. 3(b) represents the previously reported data for $Na_{0.55}CoO_2$ [5].

Figure 3(b) shows the temperature dependences of $S$ for $x = 0.65$ and 0.75 samples, together with the result reported for $Na_{1.1}Co_2O_4$ (i.e. $Na_{0.55}CoO_2$) by Kawata et al. [5]. The $S$ values for our $x = 0.65$ and 0.75 samples are ~ 75 and 120 $\mu$V / K at 300 K. These values are much larger than that reported for the $Na_{0.55}CoO_2$ sample by Kawata et al., i.e. ~ 55 $\mu$V / K at 300 K [5]. It should be noted that with increasing Na content, the $\rho$ value decreases and the $S$ value increases simultaneously. Consequently, the present Na-rich $x = 0.75$ sample shows a high value of the power-factor: $S^2 / \rho = 4.5 \times 10^{-6}$ W / cm K$^2$ (at 300 K). Having the eye on potential applications, it should be emphasized that control of the Na content in $Na_xCoO_{2+\delta}$ is most effective for the improvement of thermoelectric performance of this compound.

### 2.3 Origin of enhanced thermoelectric power

Simultaneous reduction of $\rho$ and enhancement of $S$ in $Na_xCoO_{2+\delta}$ is most peculiar. In the framework of conventional band pictures, the decrease in the $\rho$ value suggests an increase in the carrier concentration, while the increase in the $S$ value indicates the opposite (assuming that the grain boundary effect is negligible in these polycrystalline samples). The discrepancy between the dependences of $\rho$ and $S$ on the carrier concentration suggests that an unusual mechanism works for yielding the large thermoelectric power in this compound.

Recently, Koshibae et al. [11] treated thermoelectric power of cobalt oxides theoretically by generalizing the Heikes formula [13], and showed that the large magnitude of thermoelectric power observed in these compounds can be originated from both large degeneracy in spin states of cobalt species and strong correlation of the 3$d$ electrons. It was concluded that the thermoelectric power $S$ of such a compound is given by

$$S = -\frac{k_B}{e} \ln\left(\frac{g_3}{g_4} \frac{c}{1-c}\right), \quad (1)$$

where $c$, $g_3$ and $g_4$ denote the concentration of $Co^{4+}$ ions and the degrees of degeneracy for $Co^{3+}$ and $Co^{4+}$, respectively. From this formula, it is found that the thermoelectric power strongly depends on the degeneracy of cobalt species with various spin states, and the large thermoelectric power may be realized when there exists a large difference in the spin and orbital degrees of freedom between $Co^{3+}$ and $Co^{4+}$ species.

It should be noted that the magnitude of $S$ value increases with decreasing $c$, i.e. with reducing the average valence of cobalt. In $Na_xCoO_{2+\delta}$, the concentration of the $Co^{4+}$ ions, $c$, can be changed by controlling the Na content, $x$, as $c = 1-x$ when oxygen nonstoichiometry is negligible. The $S$ values calculated using Eq. (1) are summarized in Table 2. The thermoelectric power can be drastically enhanced with increasing Na content. This is consistent with our experimental finding [6]: an enhanced $S$ value was obtained in an Na-rich sample, i.e. $x = 0.75$. According to Eq. (1), the $S$ value could still be increased by further increasing Na content. Therefore, it is of great interest to enhance the solubility limit of sodium in $Na_xCoO_{2+\delta}$ to further increase the thermoelectric properties of the compound.

TABLE 2. Values of thermoelectric power calculated from Eq. (1) for $Na_xCoO_2$ with various Na contents, $x$. The $g_3 / g_4$ values in the cases (i) – (v) were taken from Ref. [11].

|  | $Co^{3+}$ | $Co^{4+}$ | $g_3 / g_4$ | $x = 0.5$ $c = 0.5$ | $x = 0.65$ $c = 0.35$ | $x = 0.75$ $c = 0.25$ |
|---|---|---|---|---|---|---|
| (i) | HS | HS | 15/6 | − 79 | − 26 | 16 |
| (ii) | HS+LS | HS | 16/6 | − 84 | − 31 | 10 |
| (iii) | LS | HS+LS | 1/12 | 214 | 267 | 309 |
| (iv) | LS | LS | 1/6 | 154 | 207 | 227 |
| (v) | HS+LS+IS | HS+LS+IS | 34/36 | 5 | 59 | 79 |

## 3. Conclusion

Here, we discussed the effects of sodium and oxygen nonstoichiometry on the thermoelectric properties of $Na_xCoO_{2+\delta}$, a promising candidate for thermoelectric applications. We established an original sample-preparation technique termed as "rapid heat-up method" that enables us to precisely control Na nonstoichiometry in $Na_xCoO_{2+\delta}$ samples by avoiding Na evaporation during the synthesis procedure. It was determined that single-phase region of the $\gamma$-$Na_xCoO_{2+\delta}$ phase is: 0.60-0.65 ≤ $x$ ≤ 0.75-0.80. This result importantly suggests that $NaCo_2O_4$ (or $Na_{0.50}CoO_2$) stoichiometry does not exist. Oxygen nonstoichiometry in our $Na_xCoO_{2+\delta}$ samples was found to be negligible from redox titration results. With increasing Na content, the $\rho$ value decreased and the $S$ value increased simultaneously. This is difficult to be understood in the framework of conventional one-electron models, suggesting an unusual mechanism for high thermoelectric power of $Na_xCoO_{2+\delta}$.

## Acknowledgments

Contributions of E. Naujalis and R. Ueda of Tokyo Institute of Technology are gratefully acknowledged. We also thank Dr. K. Isawa of Tohoku Electric Power Co. Inc. for his contribution in thermoelectric power measurements. The present work was supported by a Grant-in-aid for Scientific Research (Contract No. 11305002) from the Ministry of Education, Culture, Sports, Science and Technology of Japan.